\documentclass[intlimits,twoside,a4paper]{article}
\usepackage{amsmath,amssymb}
\usepackage{graphicx}
\usepackage{caption}
\usepackage{subcaption}
\usepackage[T2A]{fontenc}
\usepackage[cp1251]{inputenc}

\usepackage{cmpj2}
\issue{2013}{16}{2}{23006}

\doinumber{10.5488/CMP.16.23006}

\begin{document}

\title{Peculiar points in the phase diagram of the water-alcohol solutions}
\author{V.E.~Chechko\refaddr{label1}, V.Ya.~Gotsulsky\refaddr{label1}, M.P.~Malomuzh\refaddr{label2}}

\addresses{
	\addr{label1} Scientific-Research Institute of Physics, Odessa National University,
		2 Dvoryanska St., Odessa, 65026, Ukraine
	\addr{label2} Department of Theoretical Physics, Odessa National University,
		2 Dvoryasnka St., Odessa, 65026, Ukraine
}

\authorcopyright{V.E.~Chechko, V.Ya.~Gotsulsky, M.P.~Malomuzh, 2013}

\date{Received January 9, 2013, in final form April 2, 2013}

\maketitle
\begin{abstract}
The work is devoted to the investigation of nontrivial behavior of dilute
water-alcohol solutions. The tem\-pe\-ra\-tu\-re and concentration dependencies of the
contraction for aqueous solutions of ethanol and methanol are analyzed.  The
existence of a specific point, the so-called peculiar point, was established.
It is shown that water--alcohol solutions of different types obey the principle
of corresponding states if temperature and volume fraction are used as
principal coordinates.  In this case, the concentration of the peculiar point
for different solutions is close to $x_{\nu}=0.28$.  Several predictions are
made.
	
\pacs 01.55.+b
\end{abstract}

\section{Introduction}

During long time the opinion that properties of dilute molecular solutions
are close to ideal ones \cite{ref1,ref2,ref3} was widespread.  The situation
essentially changed following the first experiments on molecular light scattering (MLS)
in dilute water--alcohol solutions (see
\cite{ref4,ref5,ref6,ref7,ref8,ref9,ref10}).  It was shown that an
anomalous increase of the integral intensity for aqueous solutions of ethanol,
tertiary butanol, glycerol and others is observed if the molar concentration
approaches $x_{\textrm{M}}=0.04\div0.09$ and temperature is within some
interval characteristic of each solution. In addition, the intensity increases more
than ten times. In subsequent years, the similar anomalies were observed for
many other water-alcohol solutions.

Here, it should be noted that anomalous MLS is in fact observed at
approaching some line $L_{\mathrm{p}}$ in the plane $\left(T, x_m\right)$.  The point
$x_{\mathrm{p}}$ on this line for which the maximum of the MLS is observed was named
the peculiar point. Hereafter we will refer to the corresponding line of relative maxima as the peculiar line.

The important fact for water-glycerol solutions was established in \cite{ref11} where it
was shown that the correlation length of the concentration fluctuations
increases up to 70~\AA{} at approaching the peculiar point.  In this case, the
absolute maximum for the intensity for MLS is observed for $T_{\mathrm{p}} \approx 303$~K and
$x_{\mathrm{p}} \approx 0.04$. Such a behavior of the correlation length reflects the
existence of instability points for the concentration fluctuations.

In \cite{ref12, ref13}, the peculiar line was identified with the position of
the pseudo-spinodal for water-alcohol solutions, that separates the regions with
different cluster structures. These clusters are formed due to H-bonds
connecting water and alcohol molecules stronger than between molecules of the
same kind \cite{ref14}. In particular, according to \cite{ref12}, clusters
formed on the left of the pseudo-spinodal consist of two glycerol molecules
and ten molecules of water.  An analogous situation is also characteristic of
water-ethanol solutions \cite{ref15}.

The change of the water-alcohol structure near the peculiar line is also
manifest in the behavior of the adiabatic compressibility \cite{ref16,
ref17}, the heat capacity \cite{ref18, ref19}, diffusion of components
\cite{ref20} and other phenomena. However, all these manifestations were only
fragmentarily investigated.

In the present paper, we focus on the analysis of temperature and
concentration dependencies of the contraction for aqueous solutions of ethanol
and methanol. We want to establish the applicability of the principle of
corresponding states to the description of contraction in water-alcohol
solutions.  Some predictions will be made on this basis.

\section{Contraction and adiabatic compressibility for water-alcohol }

\begin{figure}[h]
	\centering
	\includegraphics[width=10cm]{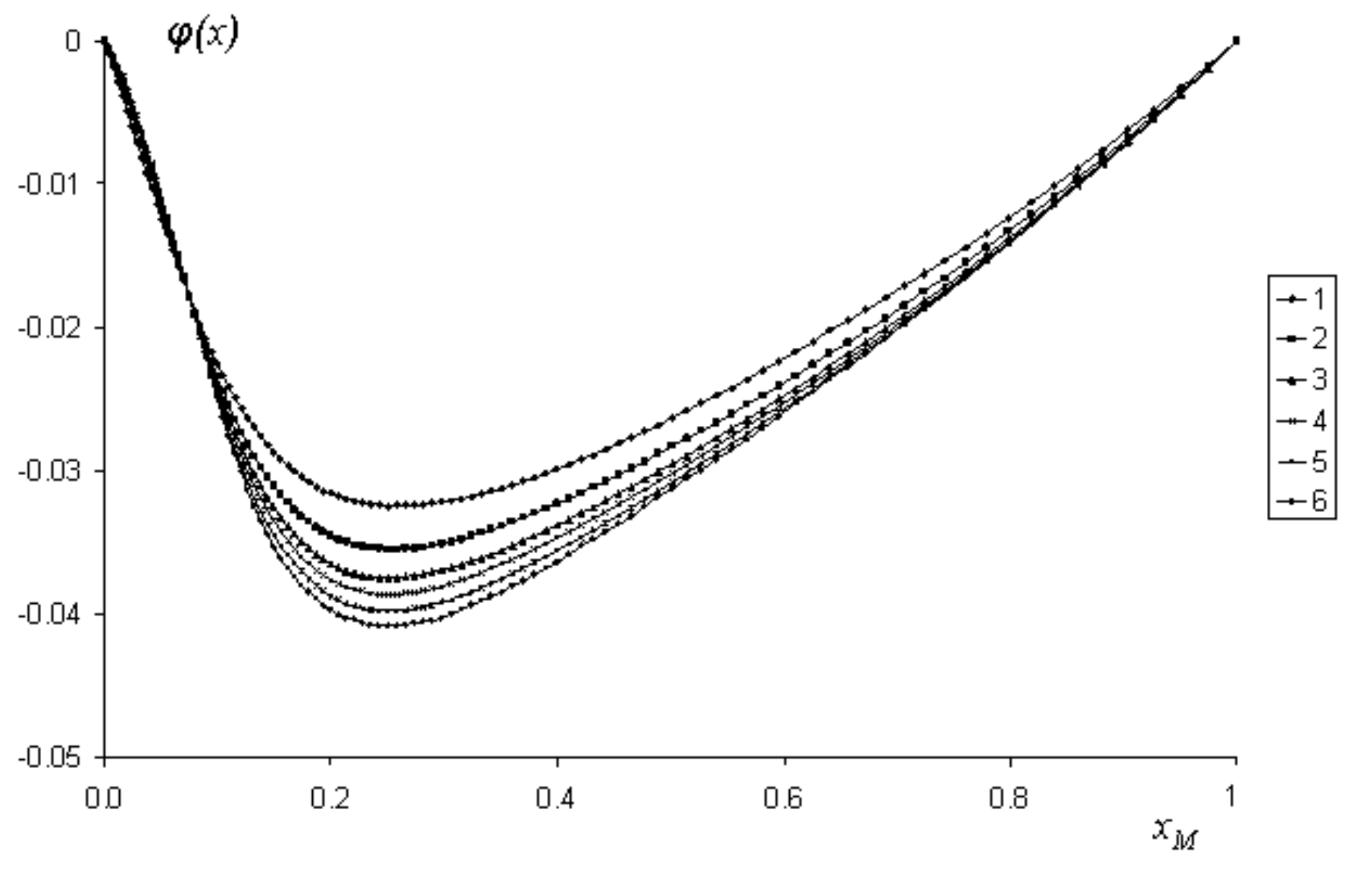}
	\caption{The concentration dependencies of $\varphi(x)$
	for water-ethanol solutions at different temperatures:
	1 --- 40$^{\circ}$C,
	2 --- 20$^{\circ}$C,
	3 --- 10$^{\circ}$C,
	4 --- 5$^{\circ}$C,
	5 --- 0$^{\circ}$C,
	6 --- --5$^{\circ}$C\,.}
	\label{fig:fig1}
\end{figure}

\begin{figure}[h]
	\centering
	\includegraphics[width=10cm]{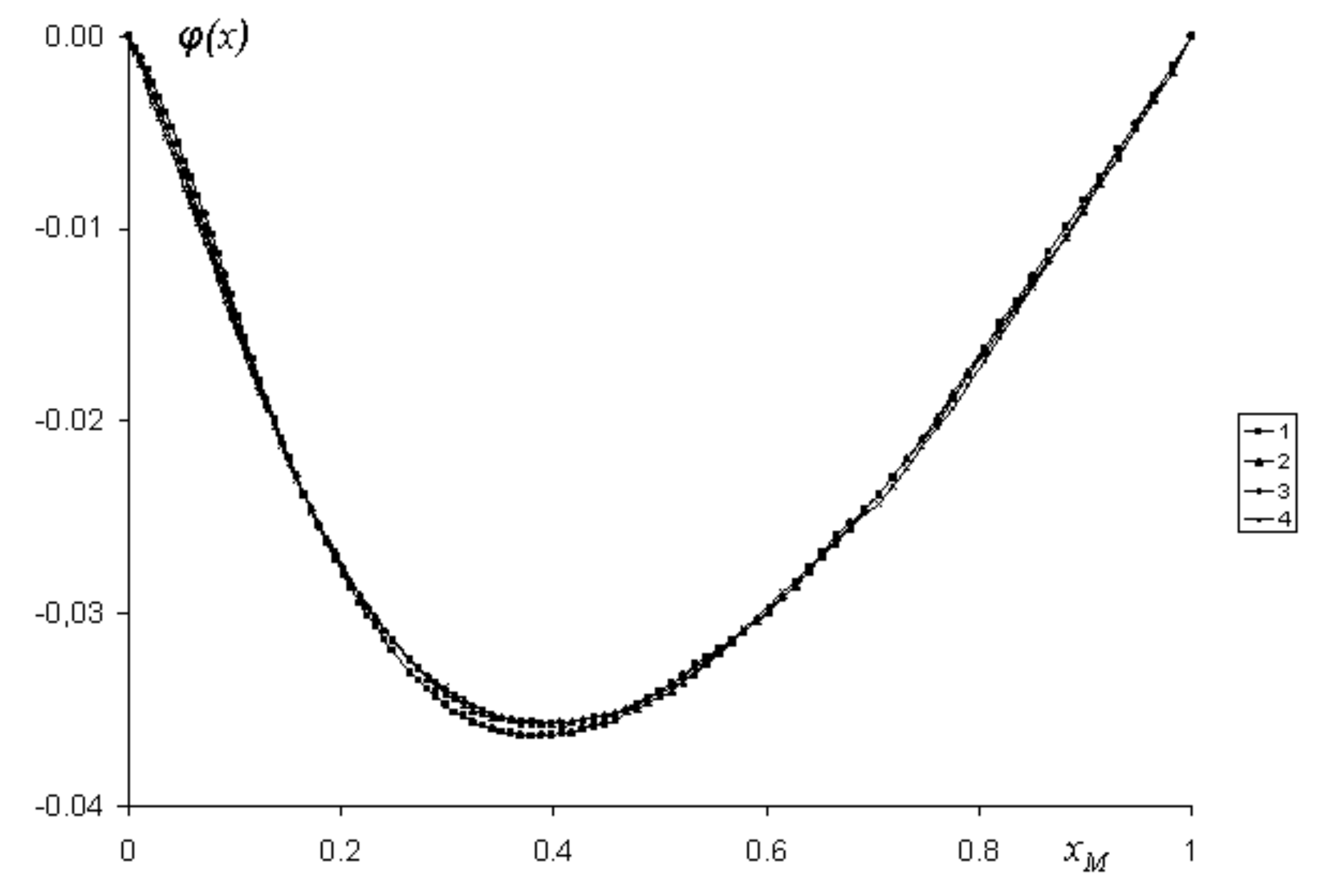}
	\caption{The concentration dependencies of $\varphi(x)$
	for water-methanol solutions at different temperatures \cite{ref23}:
		1 --- 0$^{\circ}\textrm{C}$,
		2 --- 10$^{\circ}\textrm{C}$,
		3 --- 15.6$^{\circ}\textrm{C}$,
		4 --- 20$^{\circ}\textrm{C}$\,.}
	\label{fig:fig2}
\end{figure}

By definition, the contraction of binary solution is equal to:
\[
	\varphi\left(x, T\right) = \frac{V_{12}}{V_1+V_2} - 1,
\]
where $V_i, i = 1,2$ are the initial volumes of components, $V_{12}$ is the
total volume of their mixture. It is clear that the contraction is the
simplest thermodynamic characteristic of binary solutions. Its behavior
for aqueous solutions of ethanol and methanol is presented in
figure~\ref{fig:fig1} and figure~\ref{fig:fig2}.

First, the contraction of water-ethanol solutions was investigated by Mendeleev
about 130 years ago \cite{ref21}. He established the existence of three peculiar
points: $x_\textrm{M}^{{\mathrm{(P)}}} \approx 0.077, 0.25, 0.75$, which were identified with
the intersection of straight lines used to approximate the
derivative  $\left.\dfrac{\partial \varphi}{\partial x}\right|_T$ in the concentration intervals:
$(0, 0.07), (0.1, 0.2), (0.3, 0.7), (0.8, 1)$. However, Mendeleev investigated
the behavior of the contraction only for room temperatures. Experimental data
for $V_{12}$ for other temperatures were obtained much later
\cite{ref22}. This seems very strange, but a systematic study of $\varphi\left(x\right)$
was not even carried out for typical water-alcohol solutions.

\begin{figure}[!h]
\vspace{5mm}
	\centering
	\begin{subfigure}[b]{0.48\textwidth}
	\includegraphics[width=\textwidth]{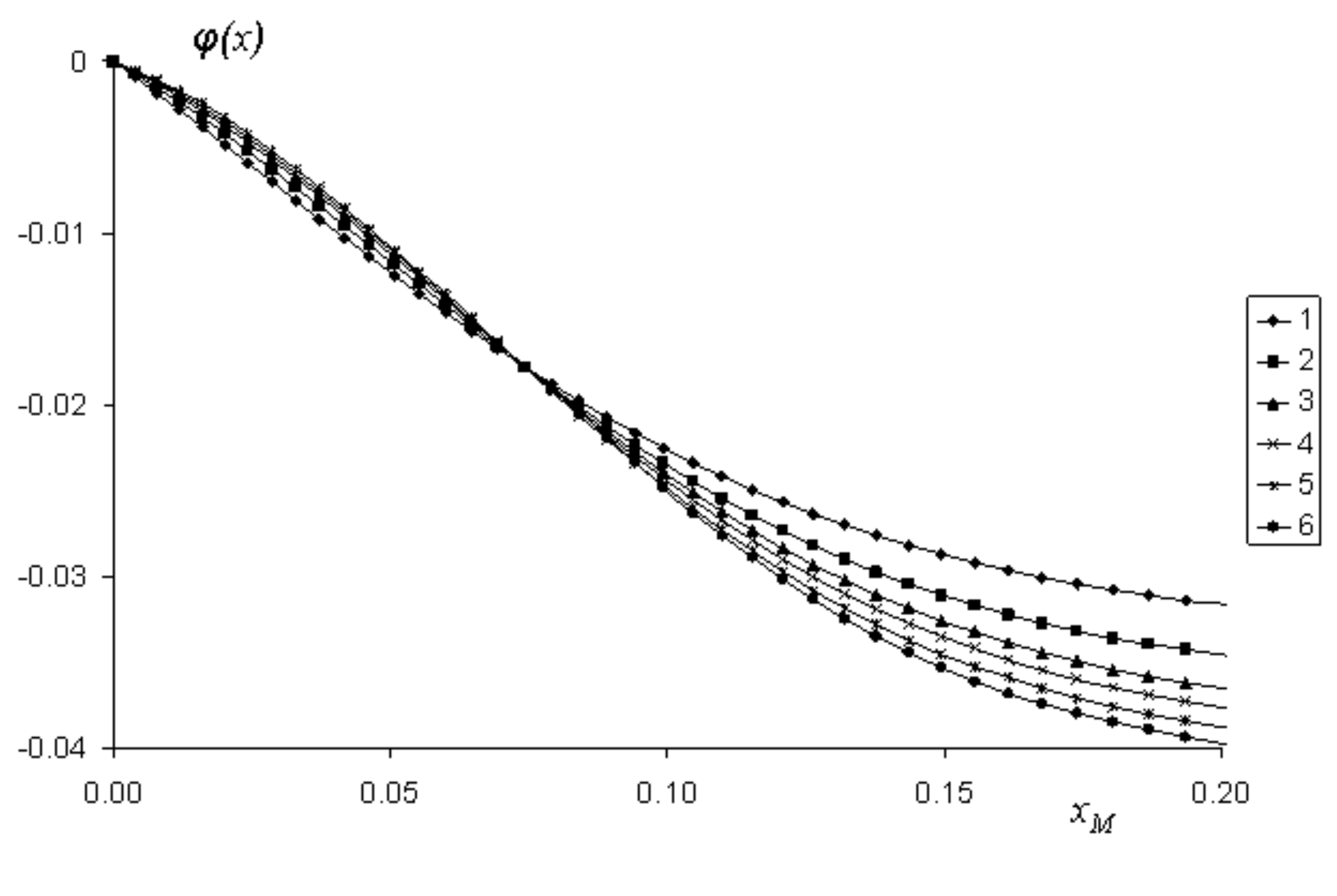}
	\caption{Ethanol}
	\label{fig:fig31}
	\end{subfigure}
	\hfill
	\begin{subfigure}[b]{0.48\textwidth}
	\includegraphics[width=\textwidth]{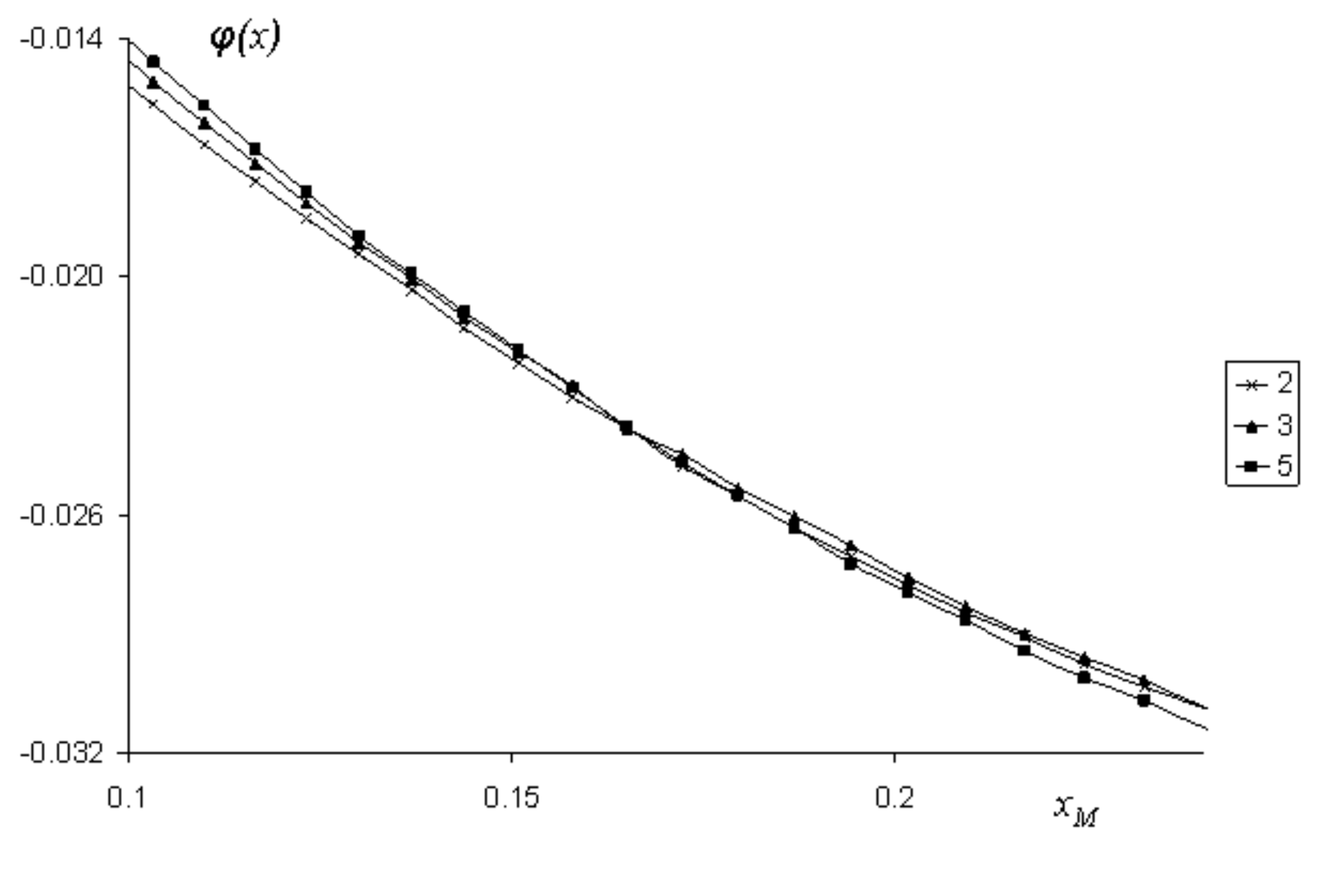}
	\caption{Methanol}
	\label{fig:fig32}
	\end{subfigure}
	
	\caption{The vicinities of peculiar points for aqueous
	solutions of ethanol and methanol at different temperatures:
		1 --- 40$^{\circ}\textrm{C}$,
		2 --- 20$^{\circ}\textrm{C}$,
		3 --- 10$^{\circ}\textrm{C}$,
		4 --- 5$^{\circ}\textrm{C}$,
		5 --- 0$^{\circ}\textrm{C}$,
		6 --- --5$^{\circ}\textrm{C}$.}
	\label{fig:fig3}
\end{figure}

As we see, the contraction curves for both aqueous solutions of ethanol and
methanol have the following peculiarities:
\begin{enumerate}
	\renewcommand{\theenumi}{\alph{enumi}}
	\renewcommand{\labelenumi}{\alph{enumi})}
	\item the signs of the contractions are negative;
	\item all curves have minima near $x_\textrm{min}^\textrm{(et)}=0.23$ and
		$x_\textrm{min}^\textrm{(met)}=0.39$;
	\item the curves corresponding to different temperatures intersect near
		$x_{\mathrm{p}}^\textrm{(et)}=0.077$ and $x_{\mathrm{p}}^\textrm{(met)}=0.12$ (see figure~\ref{fig:fig3}).
\end{enumerate}

A detailed consideration of the contraction for water-ethanol solution is
given in \cite{ref24}. Here, we will only analyze the most characteristic
properties of the contraction in the vicinity of their peculiar points in
aqueous solutions of alcohol.

The position of the peculiar point $x_{\mathrm{p}}^\textrm{(et)}=0.077$ for water-ethanol solution
practically coincides with that, registered in the MLS experiments \cite{ref25}
[$x_{\mathrm{p}}^\textrm{(et)}(\mathrm{MLS})=0.077$]. Unfortunately, there are no corresponding data for aqueous
solutions of methanol. Moreover, only for water--ethanol solutions
the existence of the peculiar point is certified with the help of contraction and the MLS,
i.e., by thermodynamic and dynamic methods.

On the other hand, the existence of the peculiar point in water-alcohol
solutions is also supported by caloric measurements. So, the specific behavior
of the heat capacity at approaching the peculiar points was observed in \cite{ref19}
for aqueous solutions of  TBA (tertiary butanol alcohol).  The heat capacity of
these solutions is characterized by sharp peaks similar to those near the
critical points. However, the height of peaks near the peculiar point remains
finite.

The strong increase of the MSL intensity near the peculiar point means that
optical homogeneity of water-alcohol solutions is violated by large scale
fluctuations, accessible for observation in the visible light range.  This
fact is immediately supported by data of correlation spectroscopy \cite{ref11},
according to which the characteristic size of heterogeneities reaches 70~\AA.
From the general point of view, an anomalous increase of fluctuations is
connected with {\it the instability of water-alcohol solutions near their peculiar
points.  However, this instability cannot be connected with delamination of
these solutions.  Usually, their delamination is observed for $0.3 <
x_\mathrm{M} < 0.5$ and sometimes for $x_\mathrm{M}$ close to $0.1$.} Therefore,
the appearance of instability should be connected with structural
transformations in water-alcohol solutions.

Experimental data on the adiabatic compressibility $\beta_\textrm{S}$ for
aqueous solutions of $\gamma$-picoline and acetone in \cite{ref16,ref17} provide
us with the important additional information on the physical nature of peculiar
points.  The quantity $\beta_\textrm{S}$ is naturally connected with the
adiabatic sound velocity: $\beta_\textrm{S} = 1
/{(\rho{c_\textrm{S}^2})}$, where $c_\textrm{S} =
\omega_\textrm{MB}/k$ and $\omega_\textrm{MB}$ is the frequency of the
Mandelshtam--Brilluoin components ($\rho$ is the mass density and $k$ is the
transfer wave vector).

All these facts allow us to conclude that (i) dilute water-alcohol solutions
can be considered as ensembles of elementary clusters. They are formed
by H-bonding between water and alcohol molecules, which is stronger
than the interaction between molecules of the same components;
(ii) elementary clusters form a percolation cluster when $x \to x_{\mathrm{p}}$.
Near $x_{\mathrm{p}}$, they begin to overlap and are destroyed; (iii) the character of
clusterization is different on the right of $x_{\mathrm{p}}$.

In other words, structural transformations in the vicinity of $x_{\mathrm{p}}$ can be
qualified as a smeared phase transition between different cluster structures.

Some details of the clusterization in aqueous solutions of glycerol--ethanol
and $\gamma$-picoline are discussed in \cite{ref12, ref13, ref15}. In these
papers it was shown that peculiar points of these solutions are located on
pseudo-spinodals separating the states of solutions with different character of
clusterization.

\section{Principle of corresponding states for the description of contraction}

The usage of molar concentration in figures~1--5 does not allow us to display
the role of molecular parameters, such as an inherent molecular volume and the
degree of nonsphericity, in the formation of the contraction. However,
exactly these parameters essentially effect the structure and size of
elementary clusters in low concentration region. In turn, this also effects
the value of molar concentration, which corresponds to the percolation
threshold.
\begin{figure}[h]
	\centering
	\label{fig:fig4}
	\includegraphics[width=8cm]{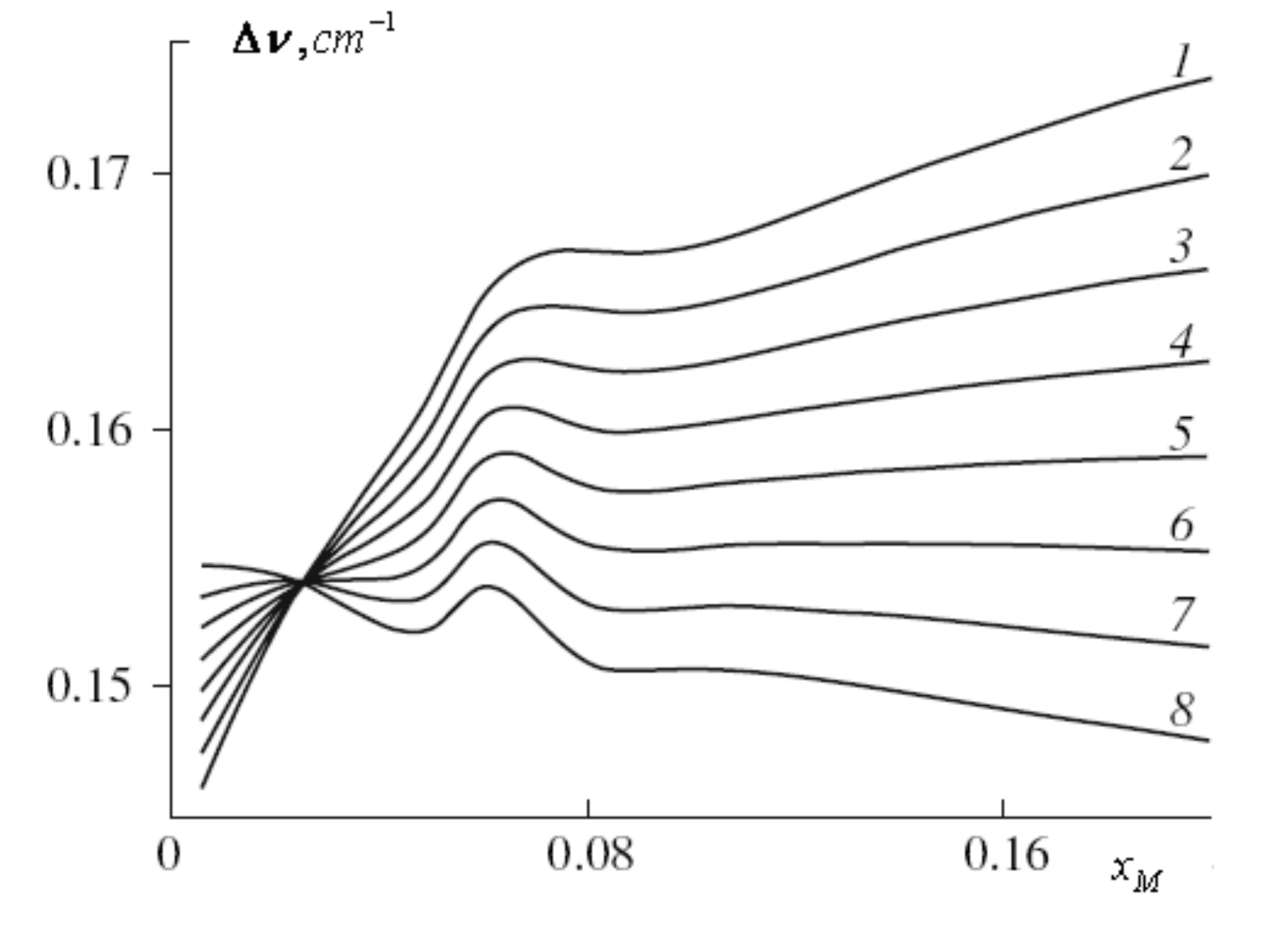}
	\caption{Isotherms of the concentration dependency for the
		Mandelshtam-Brilluoin components
		[$\Delta\nu = \omega_\textrm{MB}/\left(2\pi c\right)$, $c$ is the vacuum light velocity]
		in aqueous solutions of $\gamma$--picoline:
		1 --- 10$^{\circ}\textrm{C}$, 2 --- 20$^{\circ}\textrm{C}$, 3 --- 30$^{\circ}\textrm{C}$, 4 --- 40$^{\circ}\textrm{C}$,
		5 --- 50$^{\circ}\textrm{C}$, 6 --- 60$^{\circ}\textrm{C}$, 7 --- 70$^{\circ}\textrm{C}$, 8 --- 80$^{\circ}\textrm{C}$ \cite{ref17}.}
\end{figure}
\begin{figure}[h]
	\centering
	\label{fig:fig5}
	\includegraphics[width=7.5cm]{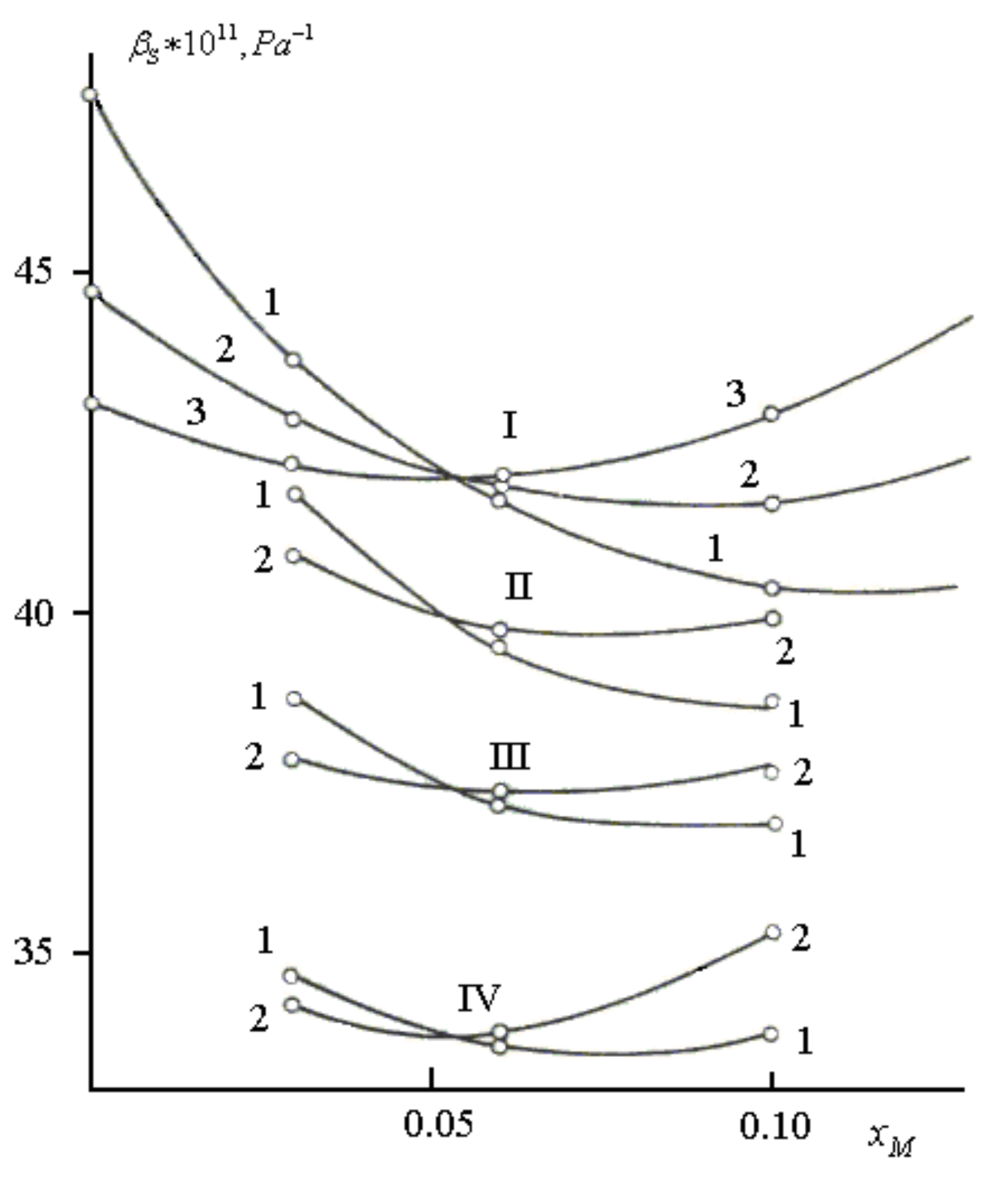}
	\caption{The adiabatic compressibility, $\beta_\textrm{S}$,
	for aqueous solutions of acetone as a function of concentration
	for different temperatures (1 --- 10$^{\circ}\textrm{C}$, 2 --- 25$^{\circ}\textrm{C}$, 3 --- 40$^{\circ}\textrm{C}$) and sodium
	chloride concentrations (I --- 0, II --- 0.4, III --- 1.0, IV --- 2.0) \cite{ref16}.
}
\end{figure}

From this point of view, it follows that elementary clusters in methanol and
ethanol should (i) be similar since they correspond to the components forming only
one H-bond between water and alcohol molecules and (ii) have different volumes,
since molecules of methanol and ethanol have noticeably different molecular
volumes.  We expect that the ratio $x_{\mathrm{p}}^\textrm{(et)} / x_{\mathrm{p}}^\textrm{(met)}$
reduces to the simplest estimate:
\begin{equation}
	\frac{x_{\mathrm{p}}^\textrm{(et)}}{x_{\mathrm{p}}^\textrm{(met)}} \Rightarrow
		\frac{\nu_m^\textrm{(met)}}{\nu_m^\textrm{(et)}} \approx 0.7.
\end{equation}
Taking the experimental values of $x_{\mathrm{p}}^\textrm{(et)}$ and $x_{\mathrm{p}}^\textrm{(met)}$,
we find $x_{\mathrm{p}}^\textrm{(et)} / x_{\mathrm{p}}^\textrm{(met)} \approx 0.72$, i.e., the
agreement between the ratios $x_{\mathrm{p}}^\textrm{(et)} / x_{\mathrm{p}}^\textrm{(met)}$ and
$\nu_m^\textrm{(met)} / \nu_m^\textrm{(et)}$ is practically full.

\begin{figure}[!h]
	\centering
	\includegraphics[width=10cm]{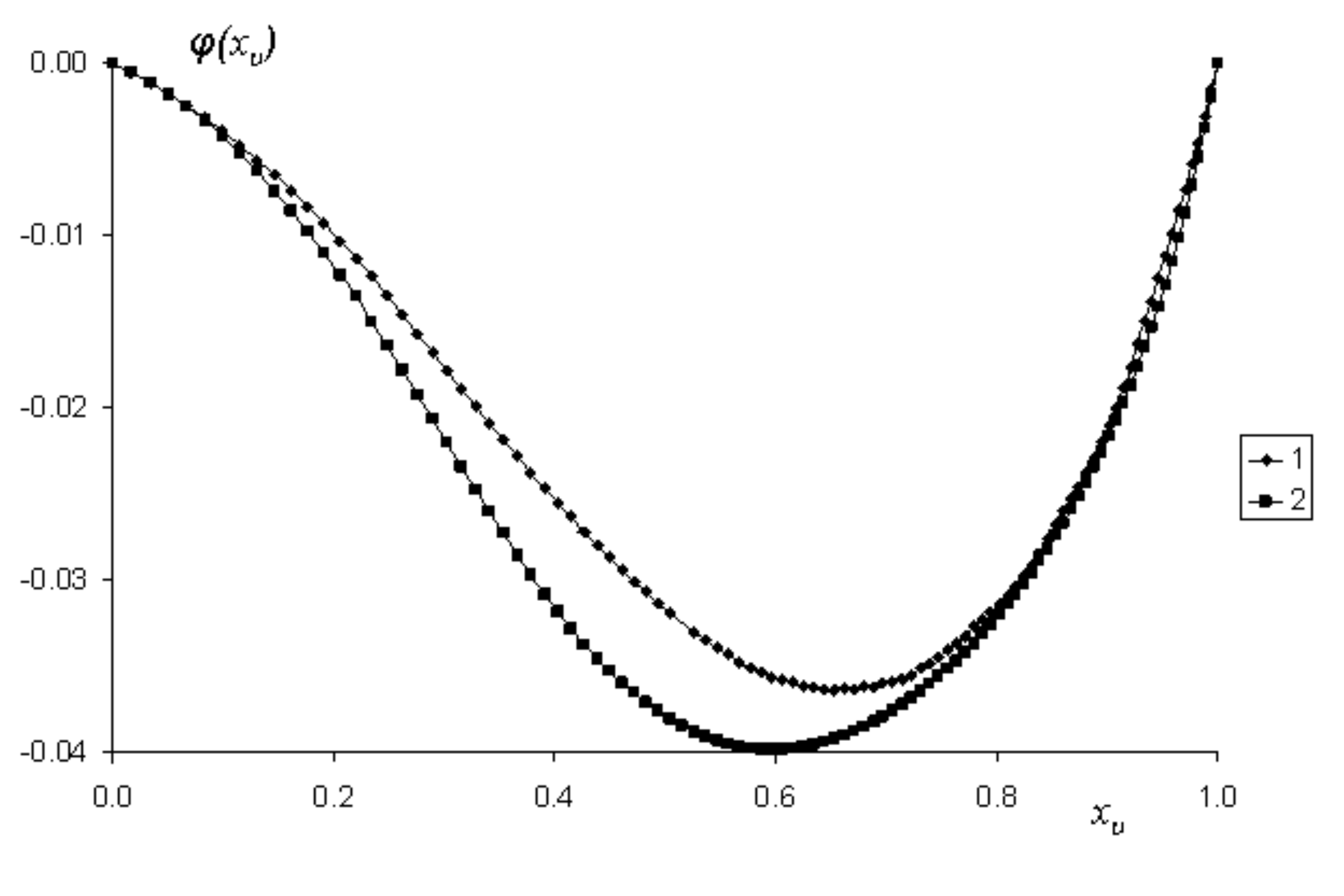}
	\caption{The contraction of water-ethanol (1)
		and water-methanol (2) solutions as a
		function of the volume fraction at $0\,^{\circ}\textrm{C}$.
	\label{fig:fig6}
}
\end{figure}

This example shows that the volume fraction $x_{\nu}$, defined as
\begin{equation}
	x_{\nu} = \frac{n_2\nu_2}{n_1\nu_1 + n_2\nu_2} \Rightarrow
		\frac{x_\textrm{M}\nu_2}{\left(1-x_\textrm{M}\right)\nu_1 + x_\textrm{M}\nu_2}~,
\end{equation}
is a more natural coordinate for the description of contraction than
the molar concentration: $x_\textrm{M} = n_2 /\left(n_1+n_2\right)$.
This conjecture is supported by figure~\ref{fig:fig6}.

For other temperatures, the agreement between methanol and ethanol curves is
analogous to that in figure~6, i.e., curves practically coincide for $0 < x < 0.2$
and $0.65 < x < 1.0$.

These facts allow us to conclude that the contraction behavior for aqueous solutions
of alcohols, for which only one H--bond forms between water and alcohol molecules, is
approximately consistent with the principle of corresponding states. In addition, the volume
fraction and temperature are natural variables for the description of the contraction behavior.
Maximal values of deviations from some universal curve do not exceed 10\%.

For low concentrations $x_{\nu} = x_\textrm{M}\nu_1/\nu_2$, and the peculiar
point is determined by the coordinate:
\begin{equation}
	x_{\nu}^{{\mathrm{(P)}}} = x_\textrm{M}^{{\mathrm{(P)}}}\frac{\nu_\textrm{a}}{\nu_\textrm{w}} + \cdots.
\end{equation}

\begin{figure}[!h]
	\centering
	\includegraphics[width=10cm]{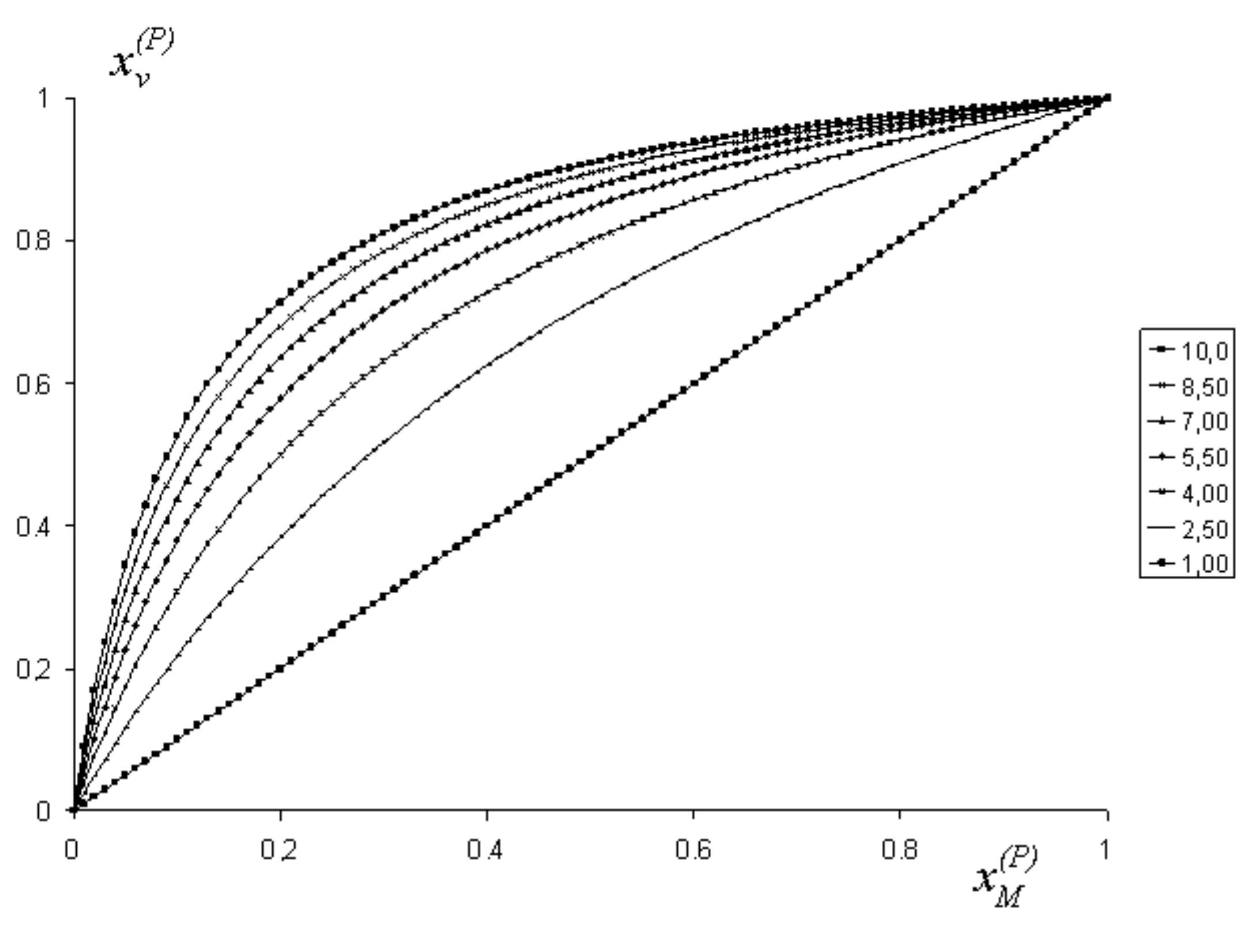}
	\caption{The interconnection between $x_{\nu}^{{\mathrm{(P)}}}$ and $x_\textrm{M}^{{\mathrm{(P)}}}$
	according to (2) for different values of the ratio $\nu_{\mathrm{w}} / \nu_{\mathrm{a}}$.
	\label{fig:fig7}
}
\end{figure}

If the ratio $\nu_{\mathrm{a}}/\nu_{\mathrm{b}}$ is much larger than unity,
$x_{\nu}^{{\mathrm{(P)}}}$ is a nonlinear function of $x_\textrm{M}^{{\mathrm{(P)}}}$
as it is demonstrated in figure~\ref{fig:fig7}.
In particular, for the ratio $x_{\nu}^{{\mathrm{(P)}}}{\mathrm{(met)}} / {x_{\nu}^{{\mathrm{(P)}}}{\mathrm{(et)}}}$, we obtain:
\[
	\frac{x_{\nu}^{{\mathrm{(P)}}}{\mathrm{(met)}}}{x_{\nu}^{{\mathrm{(P)}}}{\mathrm{(et)}}} \approx
		 \frac{x_\textrm{M}^{{\mathrm{(P)}}}{\mathrm{(met)}}}{x_\textrm{M}^{{\mathrm{(P)}}}{\mathrm{(et)}}}
		\frac{\nu_\textrm{met}}{\nu_\textrm{et}}
		 \frac{1+x_\textrm{M}^{{\mathrm{(P)}}}{\mathrm{(et)}}\left(\dfrac{\nu_\textrm{et}}{\nu_\textrm{w}}-1\right)}
		 {1+x_\textrm{M}^{{\mathrm{(P)}}}{\mathrm{(met)}}\left(\dfrac{\nu_\textrm{met}}{\nu_\textrm{w}}-1\right)}\,.
\]

\noindent From here and (1) it follows that
\begin{equation}
	\frac{x_{\nu}^{{\mathrm{(P)}}}{\mathrm{(met)}}}{x_{\nu}^{{\mathrm{(P)}}}{\mathrm{(et)}}} \approx 1.
\end{equation}
Thus, the volume fractions of methanol and ethanol at their peculiar
points are the same and we can write
\begin{equation}
	x_{\nu}^{{\mathrm{(P)}}}{\mathrm{(et)}} \approx x_{\nu}^{{\mathrm{(P)}}}{\mathrm{(met)}} \approx 0.28.
\end{equation}
In accordance with our reasoning, this is a characteristic volume fraction of
alcohol that stimulates the structural reconstruction in water-alcohol solution.
Therefore, we expect that
\begin{equation}
	x_{\nu}^{{\mathrm{(P)}}}{\mathrm{(alc)}} \approx 0.28
\end{equation}
for those water-alcohol solutions in which only one H--bond is formed between
water and alcohol mo\-le\-cu\-les.

The corresponding value for the molar concentration is equal to:
\begin{equation}
	x_{\mathrm{M}}^{{\mathrm{(P)}}}{\mathrm{(alc)}} = 0.28\frac{\nu_\textrm{w}}{\nu_\textrm{alc}}
		\frac{1}{1 - 0.28\left(1 - \dfrac{\nu_\textrm{w}}{\nu_\textrm{alc}}\right)}\,.
\end{equation}
In order to find the ratio $\nu_\textrm{w} /\nu_\textrm{alc}$, we will use the fact that
the effective molecular volume for liquids is close to the corresponding fraction volume:
\[
	\nu_i = \frac{1}{n_i} \approx \frac{m_i}{\rho_i}, \qquad i = \textrm{w}, \textrm{alc},
\]
where $n_i$ and $\rho_i$ are the number and mass densities, respectively,
and $m_i$ is the molecular mass. Then,
\begin{equation}
	\frac{\nu_\textrm{w}}{\nu_\textrm{alc}} \approx \frac{m_\textrm{w}}{m_\textrm{alc}}
		\cdot \frac{\rho_\textrm{alc}}{\rho_\textrm{w}}\,.
\end{equation}
In particular, for TBA $(\rho = 0.7887\frac{g}{{\mathrm{cm}}^3}\,,\,\, \mu = 74.12\frac{g}{{\mathrm{mol}}})$
we obtain
\begin{equation}
	x_\textrm{M}^{{\mathrm{(P)}}}(\mathrm{TBA}) \approx 0.069.
\end{equation}

\section{Discussion of the results obtained}

As was noted above, the structural transformations in water-alcohol solutions
near their peculiar points also manifest themselves in the molecular light
scattering experiments and in other phenomena. The positions of the peculiar
points, determined by different methods, are presented in table~\ref{table1}. It is taken into account that the MLS intensity usually has
maximums for two concentrations: (i) near $x_\textrm{M}{\mathrm{(alc)}} \sim 0.1 \div 0.5$,
which corresponds to the standard delamination of solution and (ii) near
$x_\textrm{M}{\mathrm{(alc)}} \sim 0.03 \div 0.08$, which corresponds to structural
transformations in the vicinity of its peculiar point. These maxima are only
observed within some temperature intervals.

\begin{table}[h]
	\centering
	\caption{Manifestations of the peculiar point in the MLS and other
	experiments.}
\vspace{2ex}
	\label{table1}
	\begin{tabular}{|p{2.8cm}|p{2.9cm}|p{2.9cm}|p{2.3cm}|p{2.0cm}|}
	\hline
	\centering Aqueous solutions \newline of some substances \newline (W-substances) &
	\centering Intensity maximum \newline in low concentrations \newline region &
	\centering Intensity maximum \newline in middle concentrations \newline region &
	\centering Inversion point \newline for the velocity \newline sound &
	\centering Minimum of the partial \newline molar volume \tabularnewline
	\hline	\hline
	W--ethanol & $0.09\,(T = 20\,\,^{\circ}\mathrm{C})$ \cite{ref25} & Indistinct \cite{ref25} &
		0.068 \cite{ref28} & 0.07 \cite{ref29} \\
	\hline
	W--methanol & $0.12\, (T = 20\,\,^{\circ}\mathrm{C})$ \cite{ref25} &  & & \\
	\hline
	W--isopropyl \newline alcohol & $0.06\, (T=20\,\,^{\circ}\mathrm{C})$ \cite{ref25}\newline $0.08$ \cite{ref8} &
		Indistinct \cite{ref25} & 0.04 \cite{ref28} & 0.06 \cite{ref29} \\
	\hline
	W--n--propyl \newline alcohol & $0.05\,(T = 20\,\,^{\circ}\mathrm{C})$ \cite{ref25} & $0.15$ \cite{ref25}
		& 0.05 \cite{ref9} & 0.045 \cite{ref29} \\
	\hline
	W--tertiary \newline butyl alcohol & $0.03$ \cite{ref25} & -- & 0.032 \cite{ref28} & -- \\
	\hline
	W--glycerol & $0.04\,(T = 10\,\,^{\circ}\mathrm{C})$ \cite{ref11} & -- & -- & -- \\
	\hline
	W--$\beta$--picolin & $0.05$ \cite{ref27} & -- & -- & -- \\
	\hline
	W--acetone & $0.055$ \cite{ref29} & $0.1$ \cite{ref9} & 0.062 \cite{ref28} & 0.07 \cite{ref29} \\
	\hline
\end{tabular}
\end{table}
Positions of the peculiar points, calculated for the same water-alcohol solutions
according to (7) and (8), are presented in table~\ref{tab:table2}.
\begin{table}[h]
	\centering
	\caption{Positions of the peculiar points determined according to (7) and (8).}
\vspace{2ex}
	\label{tab:table2}
	\begin{tabular}{|p{5cm}|p{5cm}|}
	\hline
	\centering Aqueous solutions of some substances collected in the table~1 &
	\centering Positions of the peculiar points according to (7) and (8) \tabularnewline
	\hline	\hline
	W--ethanol 			& \centering $0.09$ \tabularnewline \hline
	W--methanol 			& \centering $0.12$ \tabularnewline  \hline
	W--isopropyl alcohol 		& \centering $0.08$ \tabularnewline  \hline
	W--n--propyl alcohol 		& \centering $0.08$ \tabularnewline  \hline
	W--tertiary butyl alcohol 	& \centering $0.066$ \tabularnewline  \hline
	W--glycerol 			& \centering $0.084$ \tabularnewline  \hline
	W--$\beta$--picolin 		& \centering $0.064$ \tabularnewline  \hline
	W--acetone 			& \centering $0.083$ \tabularnewline  \hline
\end{tabular}
\end{table}

On the whole, we find a quite satisfactory agreement between experimental and
calculated data. For the majority of water-alcohol solutions, the difference
between experimental and calculated data does not exceed the width of the
interval for the experimental error. Among alcohols that form only one H-bond with
water molecules, the calculated value of $x_\textrm{M}^{{\mathrm{(P)}}}$ exceeds approximately
twice the corresponding experimental value only for tertiary butyl alcohol. A
considerable deviation of the calculated value for $x_\textrm{M}^{{\mathrm{(P)}}}$ from
experimental one is also observed for water-glycerol solution.  However, a
glycerol molecule can form three H-bonds with water molecules. Thus, it seems doubtful that  water-glycerol solutions belong to the water-ethanol similarity class.

Here, it is interesting to note that the Russian vodka and Ukrainian horilka are
aqueous solutions of ethanol having concentration of 40 volume percent.  This is more
than the concentration for the peculiar point [$x_{\nu}^{{\mathrm{(P)}}}{\mathrm{(et)}} \approx
0.28$] but it is smaller in comparison with the concentration $x_{\nu}^\textrm{(min)}{\mathrm{(et)}}
\approx 0.55$ characteristic of the minimum of contraction (see figure~\ref{fig:fig1} and
figure~\ref{fig:fig7}). Such a volume fraction of ethanol is more characteristic of
different kinds of cognacs.

It should be noted that Mendeleev was the first to relate the appearance
of peculiar points to the formation of molecular complexes in water--ethanol solutions.
Thus, he related the low concentration peculiar point at $x_\textrm{M}{\mathrm{(et)}}\approx 0.077$ for water--ethanol
solution to the formation of a complex consisting of
one molecule of ethanol and 12 molecules of water. The second peculiarity at
$x_\textrm{M}{\mathrm{(et)}} \approx 0.23$ corresponds to a complex of 1 ethanol and 3 water molecules
(see \cite{ref21}). In recent years, Mendeleev's fruitful ideas  were developed
in the works \cite{ref12,ref13,ref15,ref24,ref30,ref31}.

\bigskip
We want to dedicate this paper to 60-th anniversary of Professor Mychailo Kozlovskii
who made essential contribution to the theory of critical phenomena. We
hope that nontrivial peculiarities of dilute water--alcohol solutions will also
attract his attention.

\ukrainianpart

\title{Особливі точки фазової діаграми водно-спиртових розчинів}
\author{Чєчко В.Є.\refaddr{label1}, Гоцульський В.Я.\refaddr{label1}, Маломуж М.П.\refaddr{label2}}

\addresses{
	\addr{label1} Науково-дослідний інститут фізики, Одеський національний університет,
		Дворянська 2, Одеса, Україна
	\addr{label2} Кафедра теоретичної фізики, Одеський національний університет,
		Дворянська 2, Одеса, Україна
}

\makeukrtitle

\begin{abstract}
Робота присвячена дослідженню нетривіальної поведінки розбавлених водно-спиртових розчинів. Проаналізовано температурні і концентраційні залежності стиснення водних розчинів етанолу і метанолу. Специфічна поведінка залежностей призводить до існування особливої точки. Показано, що водно-спиртові розчини різних типів підкоряються принципу відповідних станів, якщо застосовувати в якості основних координат температуру і об'ємну частку спирту. У цьому випадку особливості різних параметрів спостерігаються в околі концентрацій близьких до $x_{\nu}=0.28$.
 Зроблено декілька прогнозів.

\keywords водно-спиртові розчини, контракція, особлива точка
\end{abstract}


\begin{thebibliography}{99}
	\bibitem{ref1}  Prigogine~I.,  The Molecular Theory of Solution. North-Holland Publishing Company Amsterdam Interscience publishers, INC., New York, 1957.

	\bibitem{ref2}  Smirnova~N.A.,  Molecular Theories of Solutions, Khimiya, Leningrad, 1987 (in Russian).

	\bibitem{ref3} Landau~L.D., Lifschitz~E.M., Statistical Physics, Part 2, Pergamon Press, Oxford, 1981.

	\bibitem{ref4}  Roshchina~G.P., In: Critical Phenomena and Fluctuations in Solutions,
			Publishing House of Acad. Sci. USSR, Moscow, 1960, 109--116 (in Russian).

	\bibitem{ref5} Eskin V.E., Nesterov A.E., Ukr. Fiz. Zh., 1964, \textbf{9}, No.~2, 540--544 (in Russian).
	
	\bibitem{ref6}  Beer~C.W.,   Jolly~D.J., Opt. Commun., 1974, \textbf{11}, Iss.~2, 150--151; \doi{10.1016/0030-4018(74)90205-3}.

	\bibitem{ref7} Vuks~M.F., Light Scattering in Gases, Liquids and Solutions,
		Leningrag. University Press, Leningrad, 1977 (in Russian).

	\bibitem{ref8} Lanshina L.V., Russ. J. Phys. Chem. A, 1998, \textbf{72}, No.~7, 1110.
	
	\bibitem{ref9} Subramanian D., Anisimov M.A., J.~Phys. Chem. B, 2011, \textbf{115}, Iss.~29, 9179--9183; \doi{10.1021/jp2041795}.

	\bibitem{ref10} Chechko~V.Eu., Lokotosh T. V., Malomuzh N. P., Zaremba V. G., Gotsulsky~V.Ya., J. Phys. Stud., 2003, \textbf{7}, No.~2, 175--183.

	\bibitem{ref11} Chechko V. E., Gotsulskiy V. Ya., Zaremba V.G., J. Mol. Liq., 2003, \textbf{105}, Iss.~2--3, 211--214; \doi{10.1016/S0167-7322(03)00055-2}.

	\bibitem{ref12} Malomuzh~N.P., Slinchak~E.L., Russ. J. Phys. Chem.~A, 2007, \textbf{81}, No.~11, 1777--1782; \doi{10.1134/S0036024407110106}.

	\bibitem{ref13} Malomuzh N.P., Slinchak E.L., Ukr. J. Phys., 2008, \textbf{53}, No.~10, 966--970.
	
	\bibitem{ref14} Pimentel G.C., McClellan A.L., The Hydrogen Bond, W.H. Freeman Publishers, San Francisco, 1960.
	
	\bibitem{ref15} Malomuzh N.P., Pankratov K.N., Slinchak E.L., Ukr. J. Phys. 2008, \textbf{53}, No.~11, 1080--1085.
	
	\bibitem{ref17} Dakar G.M., Khakimov P.A, Korikova M.L., J. Phys. Chem., 1992, \textbf{66}, No.~1, 200 (in Russian).
	
	\bibitem{ref16}  Sabirov~L.M.,   Semenov~D.I., Kha\u{\i}darov~Kh.S., Opt. Spectrosc., 2008, \textbf{105}, No.~3, 369--376; \doi{10.1134/S0030400X08090087}.
	
	\bibitem{ref18} Anisimov M.A., Esipov V.S.,  Zaprudskii~V.M. , Zaugol'nikova~N.S., Ovodov~G.I., Ovodova~T.M., Seifer~A.L., J. Struct. Chem., 1977, \textbf{18}, No.~5, 663--670; \doi{10.1007/BF00746104}.
	
	\bibitem{ref19} Anisimov~M.A., Critical Phenomena in Liquids and Liquid Crystals, Nauka, Moscow, 1987 (in Russian).
	
	\bibitem{ref20}  Bulavin~L.,   Slisenko~V.,   Vasylkevych~O.,
		 Kovalyov~O.,   Krotenko~V.,  Korbetskyi~E., In: Modern Problems of Molecular Physics, Kyiv National University Publ. House, Kyiv, 2006, 73--78 (in Ukrainian).
	
	\bibitem{ref21} Mendeleev D.I., Solutions. The series ``Classics of Science'',
		Publishing House of Acad. Sci. USSR, Moscow, 1956 (in Russian).
	
	\bibitem{ref22} Tables for the determination of ethanol in water-alcohol solution. State Committee of Standards Council of Ministers, Moscow, 1972 (in Russian).
	
	\bibitem{ref23} CRC Handbook of Chemistry and Physics, 44th ed., 1962, 2582--2584.
	
	\bibitem{ref24} Gotsulskiy~V.Ya., Malomuzh~N.P., Chechko~V.E.,
		Russ. J. Phys. Chem. A (unpublished).
	
	\bibitem{ref25} Vuks M.F., Shurupova L.V.,
		Opt. Spectrosc., 1976, \textbf{40}, No.~1, 154--159 (in Russian).
	
	\bibitem{ref26} Eskin V.E., Nesterov A.E.,
		Ukr. Fiz. Zh., 1964, \textbf{9}, No.~2, 540 (in Russian).
	
	\bibitem{ref27} Endo H., Bull. Chem. Soc. Jpn., 1973, \textbf{46}, No.~4., 1106--1111; \doi{10.1246/bcsj.46.1106}.
	
	\bibitem{ref28} Chaban I.A., Sov. Phys. Acoust., 1975, \textbf{21}, No.~2, 286--293 (in Russian).
	
	\bibitem{ref29} Tonaka H., Nakanishi K., Touhara H., J. Chem. Phys., 1984, \textbf{81}, 4065--4073; \doi{10.1063/1.448150}.
	
	\bibitem{ref30} Atamas A.A., Atamas N.A., Bulavin L.A., J. Mol. Liq., 2005, \textbf{120}, 15--17; \doi{10.1016/j.molliq.2004.07.073}.
	
	\bibitem{ref31} Atamas A.A., Atamas N.A., Bulavin L.A., Russ. J. Phys. Chem.~A, 2005,
		 \textbf{79}, No.~8, 1428--1432 (in Russian).

\end{thebibliography}
\end{document}